\documentclass[12pt]{article}
\usepackage{amsfonts,amsmath,amssymb}
\usepackage{theorem}
\theorembodyfont{\rmfamily}

\newcommand{\CC}{{\mathbb C}}

\newcommand{\ZZ}{{\mathbb Z}}

\def\lddots{\mathinner{\mkern1mu\raise1pt\hbox{.}\mkern2mu
\raise4pt\hbox{.}\mkern2mu\raise7pt\vbox{\kern7pt\hbox{.}}\mkern1mu}}

\newcommand{\non}{\nonumber}
\newcommand{\tr}{\mathop{\rm Tr}\nolimits}

\def\qmbox#1{\qquad\mbox{#1}\quad}

\begin{document}


\strut\hfill{}

\vspace{0.5in}

\begin{center}

{\Large \textsf{Bethe Ansatz equations and exact $S$ matrices \\[2mm]
for the $osp(M|2n)$ open super spin chain.\\[5mm]
}}

\vspace{10mm}

{\large D. Arnaudon$^a$, J. Avan$^b$ , N.~Cramp\'e$^a$,
A.~Doikou$^a$, L. Frappat$^{ac}$, {E}. Ragoucy$^a$}

\vspace{10mm}

\emph{$^a$ Laboratoire d'Annecy-le-Vieux de Physique
Th{\'e}orique}

\emph{LAPTH, CNRS, UMR 5108, Universit{\'e} de Savoie}

\emph{B.P. 110, F-74941 Annecy-le-Vieux Cedex, France}

\vspace{7mm}

\emph{$^b$ Laboratoire de Physique Th{\'e}orique et
Mod\'elisation, CNRS UMR 8089}

\emph{Universit\'e de Cergy, 5 mail Gay-Lussac, Neuville-sur-Oise}

\emph{F-95031 Cergy-Pontoise Cedex}

\vspace{7mm}

\emph{$^c$ Member of Institut Universitaire de France}

\end{center}

\vfill \vfill

\begin{abstract}
We formulate the Bethe Ansatz equations for the open
super spin chain based on the super Yangian of $osp(M|2n)$ and 
with diagonal boundary conditions. We then study
the bulk and boundary scattering of the $osp(1|2n)$ open spin
chain.
\end{abstract}

\vfill MSC number: 81R50, 17B37 \vfill

\rightline{LAPTH-1000/03} \rightline{math-ph/0310042}
\rightline{October 2003}

\baselineskip=16pt

\newpage

\section{Introduction}
The notion of the reflection equation associated with solutions of
the Yang--Baxter equation \cite{baxter, korepin}, goes back to the
key works of Cherednik \cite{cherednik} and Sklyanin
\cite{sklyanin}. The subject has recently attracted a great deal
of activity as was summarised in \cite{yabon} (and references
therein). More specifically, starting from a quantum $R$-matrix
$R(\lambda)$ depending on the spectral parameter $\lambda$ and
satisfying the ({\it super}) Yang--Baxter equation \cite{baxter,
korepin, kulish}
\begin{equation}
  R_{12}(\lambda_{1} - \lambda_{2})\ R_{13}(\lambda_{1})\ R_{23}(\lambda_{2})
  = R_{23}(\lambda_{2})\ R_{13}(\lambda_{1})\ R_{12}(\lambda_{1} -
\lambda_{2}) \,, \label{YBE}
\end{equation}
one derives the reflection equation for an object $K(\lambda)$ as
\begin{equation}
  R_{12}(\lambda_1-\lambda_2)\ K_{1}(\lambda_1)\
  R_{12}(\lambda_1+\lambda_2)\ K_{2}(\lambda_2)=
  K_{2}(\lambda_2)\ R_{12}(\lambda_1+\lambda_2)\
  K_{1}(\lambda_1)\ R_{12}(\lambda_1-\lambda_2)\,.
  \label{reflection}
\end{equation}
We have proposed in \cite{yabon} a classification of $c$-number
solutions $K(\lambda)$ of the reflection equation
(\ref{reflection}) for rational (super) Yangian $R$-matrices
\cite{Dr, soya} associated to the infinite series $so(m)$,
$sp(2n)$ and $osp(m|2n)$. This classification entailed $K$ matrices
with purely diagonal, anti-diagonal and mixed (diagonal, 
anti-diagonal) non-zero entries. The explicit values of the $K$
matrices were then used within the analytical Bethe Ansatz
formulation \cite{reshe, mnanal, mnanal2, doikou, yabon} for the
derivation of the spectrum and the bulk and boundary $S$-matrices
for the $so(m)$, $sp(2n)$ open spin chains.
\\
There exists a substantial body of work on $gl(m|n)$ super spin
chains.
Interest in these systems stemmed from the existence of 
physically relevant particular cases such as
supersymmetric $t$-$J$ and extended Hubbard models.
They have been the object of many studies.
Supersymmetric $t$-$J$ models were considered, e.g. in \cite{schlot}
(thermodynamical aspects),
\cite{GR94} (diagonal boundary $K$ matrices) and
\cite{essler2} (boundary $S$-matrix).
Extended Hubbard models were considered in \cite{essler1} (closed
chain) and in \cite{ENSLAPP650,zhazhou,gufogros} (open chains with
integrable boundary conditions), 
whilst spin ladder systems associated to some $sl(m|n)$ superalgebras 
were obtained in \cite{fohiliro,tofohili}. 
General results for continuum limit of the $gl(m|n)$ super spin chains
were derived in \cite{saleur1}.
\\
A natural alternative to these models with $gl(m|n)$ underlying
superalgebras is provided by super spin chains with underlying
$osp(m|2n)$ superalgebras. 
A connection to intersecting loop models and hence polymer field
theories was pointed out in \cite{maniri}, where the analytical Bethe
Ansatz equations were written for the closed spin chain. 
Algebraic methods were used for some specific cases in
\cite{ENSLAPP505,mara} including nested Bethe Ansatz in
\cite{mara}. 
Field theoretical limits were also considered in the literature: 
the exact bulk $osp(2|2)$ $S$-matrix was conjectured in
\cite{bale} in the framework of disordered systems. 
Investigation of the thermodynamics of $osp(1|2n)$ closed spin chains
was undertaken in \cite{tsuboi} using the thermodynamical Bethe Ansatz
formalism. An algebraic construction using Birman--Wenzl--Murakami
algebra then yielded conjectural $S$-matrices for 
field theoretical $osp(m|2n)$ models, and allowed a subsequent
thermodynamical 
Bethe Ansatz  analysis of their thermodynamical properties
\cite{saweka}.  
However a systematic thermodynamic
treatment of these models with more general boundaries is still
missing.
\\
Our purpose is to make an exhaustive study of the more
complicated case of open spin chains with $osp(m|2n)$ underlying
superalgebra and any integrable (diagonal at a first step) boundary
conditions. The strategy is to establish (insofar as the methods are
available) Bethe Ansatz equations for ground state and excited states
(note that there is no obvious relation between closed spin chains
Bethe Ansatz equations and open spin chain Bethe Ansatz equations,
particularly when non trivial boundary conditions are involved); solve
them within the non trivial string hypothesis (discussed in the closed
case in \cite{schlot,saleur1}); and use the results to obtain  the
$S$-matrix and thermodynamical quantities, with explicit evaluation of
the effect of boundary conditions.

This paper is our first step in this direction: using the analytical
Bethe Ansatz method, we derive the
Bethe Ansatz equations for all orthosymplectic superalgebras, and
all \emph{diagonal} $K$ matrices. Restricting ourselves then to
$osp(1|2n)$, we solve these equations in the thermodynamic limit,
we derive the ground state and low-lying excitations, and compute
explicitly the bulk and boundary $S$-matrices. Further
generalisations will be left for future investigations.

\section{Bethe Ansatz equations for the $\mathbf{osp(M|2n)}$ open spin
chain} \setcounter{equation}{0}
\subsection{Conventions and notations}
The Bethe Ansatz equations will be derived here for the
$osp(M|2n)$ $N$-site open spin chain with diagonal reflection
conditions by means of the analytical Bethe Ansatz method (see
e.g. \cite{reshe,mnanal,doikou}). As customary to construct the
open chain transfer matrix we introduce the $R$-matrix which is a
solution of the {\it super} Yang--Baxter equation. We focus on the
$osp(M|2n)$ invariant $R$-matrix given by \cite{soya}
\begin{eqnarray}
  R(\lambda) = \lambda(\lambda+i\kappa)1 +i(\lambda +i\kappa) P -i\lambda Q\;,
  \qquad 2\kappa=\theta_0(M-2n-2)\label{r}
\end{eqnarray}
where $P$ is the (super)permutation operator (i.e.  $X_{21}\equiv
PX_{12}P$)
\begin{equation}
  \label{eq:Pdef}
  P = \sum_{i,j=1}^{M+2n} (-1)^{[j]} E_{ij} \otimes E_{ji}
\end{equation}
and
\begin{equation}
  \label{eq:Qdef}
  Q = \sum_{i,j=1}^{M+2n} (-1)^{[i][j]} \theta_{i} \theta_{j}
  E_{\bar{\jmath}\bar{\imath}} \otimes E_{ji} \equiv P^{t_{1}} \;.
\end{equation}
For each index $i$, we have introduced a conjugate index
\begin{equation}
\bar\imath=M+2n+1-i\,.
\end{equation}
We also introduce a sign $\theta_i$ and a $\ZZ_2$-grading $[i]$
whose definition, due to the conventions we adopt (see below),
depend whether we consider the superalgebra $osp(2|2n)$ or any
other $osp(M|2n)$ superalgebra:
\\
{\bf For $osp(M|2n)$ superalgebras, $M\neq2$:}
\begin{equation}
  \label{eq:deftheta}
  \theta_i = \begin{cases}
    +1 & \qmbox{for} 1\le i\le M+n  \\
    -1 & \qmbox{for} M+n+1 \le i\le M+2n
  \end{cases}
\end{equation}
\begin{alignat}{3}
  \label{eq:defigrad}
    &(-1)^{[i]} = +1 &&
    \qmbox{for} 1\le i\le n\qmbox{and} M+1\le i\le M+2n\\
    &(-1)^{[i]} = -1
    && \qmbox{for} n+1\le i \le n+M
\end{alignat}
We will associate to this choice the sign $\theta_0=-1$.\\
{\bf For $osp(2|2n)$ superalgebras:}
\begin{equation}
  \label{eq:deftheta2}
  \theta_i = \begin{cases}
    +1 & \qmbox{for} 1 \le i\le n+1\qmbox{and} i=2n+2\\
    -1 & \qmbox{for} n+2 \le i\le 2n+1
  \end{cases}
\end{equation}
\begin{alignat}{3}
  \label{eq:defigrad2}
    &(-1)^{[i]} = +1
    && \qmbox{for} i=1\qmbox{and} i=2n+2\\
    &(-1)^{[i]} = -1 &&
    \qmbox{for} 2\le i\le 2n+1
\end{alignat}
The sign corresponding to this choice will be $\theta_0=+1$.

\medskip

The transposition $^t$ used in (\ref{eq:Qdef}) and below is
defined, for $A = \sum_{ij} \; A^{ij} \;E_{ij}$, by
\begin{equation}
  \label{eq:t}
  A^t = \sum_{ij} (-1)^{[i][j]+[j]} \theta_i \theta_j \; A^{ij} \,
  E_{\bar\jmath \bar\imath}
  = \sum_{ij}  \left(A^{t}\right)^{ij} \,
  E_{ij}
\end{equation}
The $R$-matrix (\ref{r}) satisfies crossing and unitarity, namely
\begin{eqnarray}
  R_{12}(\lambda) R_{12}(-\lambda)  =
  (\lambda^2+\kappa^2)(\lambda^2+1)\, 1, ~~R_{12}(\lambda) =
  R_{12}^{t_{1}}(-\lambda -i\kappa)\,. \label{cu}
\end{eqnarray}
We finally define the super trace operation according to the
$\ZZ_2$-grading we have introduced:
\begin{equation}
\tr A = \sum_{j=1}^{M+2n} (-1)^{[j]}\, A_{jj} \mbox{ for }
A=\sum_{i,j=1}^{M+2n} A_{ij}\, E_{ij}\,. \label{def-supertrace}
\end{equation}
\subsection{Transfer matrix and pseudo vacuum}
The open chain transfer matrix is given by \cite{sklyanin}
\begin{eqnarray}
  t(\lambda) = \tr_{0} K_{0}^{+}(\lambda)\ T_{0}(\lambda)\
  K^{-}_{0}(\lambda)\ \hat T_{ 0}(\lambda)\,, \label{transfer1}
\end{eqnarray} where $\tr_{0}$ denotes here the {\it super} trace
(\ref{def-supertrace}) over the auxiliary space,
\begin{eqnarray}
  T_{0}(\lambda) = R_{0N}(\lambda) R_{0,N-1}(\lambda) \cdots R_{0
    2}(\lambda) R_{01}(\lambda)\,,~~\hat T_{0}(\lambda) =
  R_{10}(\lambda) R_{20}(\lambda) \cdots R_{N-1,0}(\lambda)
  R_{N0}(\lambda)\,, \label{hatmonodromy}
\end{eqnarray}
$K^{-}_{0}(\lambda)$ is any solution of the {\it
  super}
boundary Yang--Baxter equation
\begin{equation}
  R_{12}(\lambda_1-\lambda_2)\, K_{1}(\lambda_1)\,
  R_{12}(\lambda_1+\lambda_2)\, K_{2}(\lambda_2) =
  K_{2}(\lambda_2)\, R_{12}(\lambda_1+\lambda_2)\,
  K_{1}(\lambda_1)\, R_{12}(\lambda_1-\lambda_2)
  \label{eq:bybe}
\end{equation}
and $K^{+}_{0}(\lambda)$ is a solution of a closely related
reflection equation defined to be:
\begin{equation}
  R_{12}(\lambda_2-\lambda_1)\, K_{1}^{t_{1}}(\lambda_1)\, R_{12}(-\lambda_1-\lambda_2-2i\kappa)\, K_{2}^{t_{2}}(\lambda_2) =
  K_{2}^{t_{2}}(\lambda_2)\, R_{12}(-\lambda_1-\lambda_2-2i\kappa)\, K_{1}^{t_{1}}(\lambda_1)\, R_{12}(\lambda_2-\lambda_1)\,.
  \label{eq:bybeplus}
\end{equation}
It is clear that any solution $K^{-}(\lambda)$ of (\ref{eq:bybe}),
e.g. given in \cite{yabon},  gives rise to a solution
$K^{+}(\lambda)$ of (\ref{eq:bybeplus}),  defined by
$K^{+}(\lambda) = K^{-}(-\lambda -i\kappa)^{t}$.

\begin{figure}[ht]
\bigskip
\begin{center}
$osp(2m|2n)$, $m>1$
\end{center}
\begin{center}
\begin{picture}(260,40) \thicklines
\multiput(0,20)(42,0){5}{\circle{14}}
\put(0,35){\makebox(0,0){$a_1$}}
\put(42,35){\makebox(0,0){$a_{n-1}$}}
\put(84,35){\makebox(0,0){$a_n$}}
\put(126,35){\makebox(0,0){$a_{n+1}$}}
\put(168,35){\makebox(0,0){$a_{n+m-2}$}}
\put(79,15){\line(1,1){10}}\put(79,25){\line(1,-1){10}}
\put(7,20){\line(1,0){4}}\put(15,20){\line(1,0){4}}
\put(23,20){\line(1,0){4}}\put(31,20){\line(1,0){4}}
\put(49,20){\line(1,0){28}} \put(91,20){\line(1,0){28}}
\put(133,20){\line(1,0){4}}\put(141,20){\line(1,0){4}}
\put(149,20){\line(1,0){4}}\put(157,20){\line(1,0){4}}
\put(173,25){\line(2,1){20}}\put(173,15){\line(2,-1){20}}
\put(199,40){\circle{14}} \put(230,40){\makebox(0,0){$a_{n+m-1}$}}
\put(199,0){\circle{14}} \put(230,0){\makebox(0,0){$a_{n+m}$}}
\end{picture}
\end{center}

\bigskip

\begin{center}
$osp(2m+1|2n)$, $m\geq1$
\end{center}
\begin{center}
\begin{picture}(220,20) \thicklines
\multiput(0,0)(42,0){6}{\circle{14}}
\put(0,15){\makebox(0,0){$a_1$}}
\put(42,15){\makebox(0,0){$a_{n-1}$}}
\put(84,15){\makebox(0,0){$a_n$}}
\put(126,15){\makebox(0,0){$a_{n+1}$}}
\put(168,15){\makebox(0,0){$a_{n+m-1}$}}
\put(210,15){\makebox(0,0){$a_{n+m}$}}
\put(79,-5){\line(1,1){10}}\put(79,5){\line(1,-1){10}}
\put(7,0){\line(1,0){4}}\put(15,0){\line(1,0){4}}
\put(23,0){\line(1,0){4}}\put(31,0){\line(1,0){4}}
\put(49,0){\line(1,0){28}} \put(91,0){\line(1,0){28}}
\put(133,0){\line(1,0){4}}\put(141,0){\line(1,0){4}}
\put(149,0){\line(1,0){4}}\put(157,0){\line(1,0){4}}
\put(174,-3){\line(1,0){30}}\put(174,3){\line(1,0){30}}
\put(195,0){\line(-1,1){10}}\put(195,0){\line(-1,-1){10}}
\end{picture}
\end{center}

\bigskip

\begin{center}
$osp(1|2n)$
\end{center}
\begin{center}
\begin{picture}(100,20) \thicklines
\put(0,0){\circle{14}} \put(42,0){\circle{14}}
\put(84,0){\circle*{14}} \put(0,15){\makebox(0,0){$a_1$}}
\put(42,15){\makebox(0,0){$a_{n-1}$}}
\put(84,15){\makebox(0,0){$a_n$}}
\put(7,0){\line(1,0){4}}\put(15,0){\line(1,0){4}}
\put(23,0){\line(1,0){4}}\put(31,0){\line(1,0){4}}
\put(48,-3){\line(1,0){30}}\put(48,3){\line(1,0){30}}
\put(69,0){\line(-1,1){10}}\put(69,0){\line(-1,-1){10}}
\end{picture}
\end{center}

\bigskip

\begin{center}
$osp(2|2n)$
\end{center}
\begin{center}
\begin{picture}(140,20) \thicklines
\multiput(0,0)(42,0){4}{\circle{14}}
\put(0,15){\makebox(0,0){$a_1$}} \put(42,15){\makebox(0,0){$a_2$}}
\put(84,15){\makebox(0,0){$a_n$}}
\put(126,15){\makebox(0,0){$a_{n+1}$}}
\put(-5,-5){\line(1,1){10}}\put(-5,5){\line(1,-1){10}}
\put(7,0){\line(1,0){28}}
\put(49,0){\line(1,0){4}}\put(57,0){\line(1,0){4}}
\put(65,0){\line(1,0){4}}\put(73,0){\line(1,0){4}}
\put(101,0){\line(1,1){10}}\put(101,0){\line(1,-1){10}}
\put(90,-3){\line(1,0){30}}\put(90,3){\line(1,0){30}}
\end{picture}
\end{center}
\caption{Distinguished Dynkin diagrams of the $osp(M|2n)$
superalgebras. \label{fig:DD1}}
\end{figure}
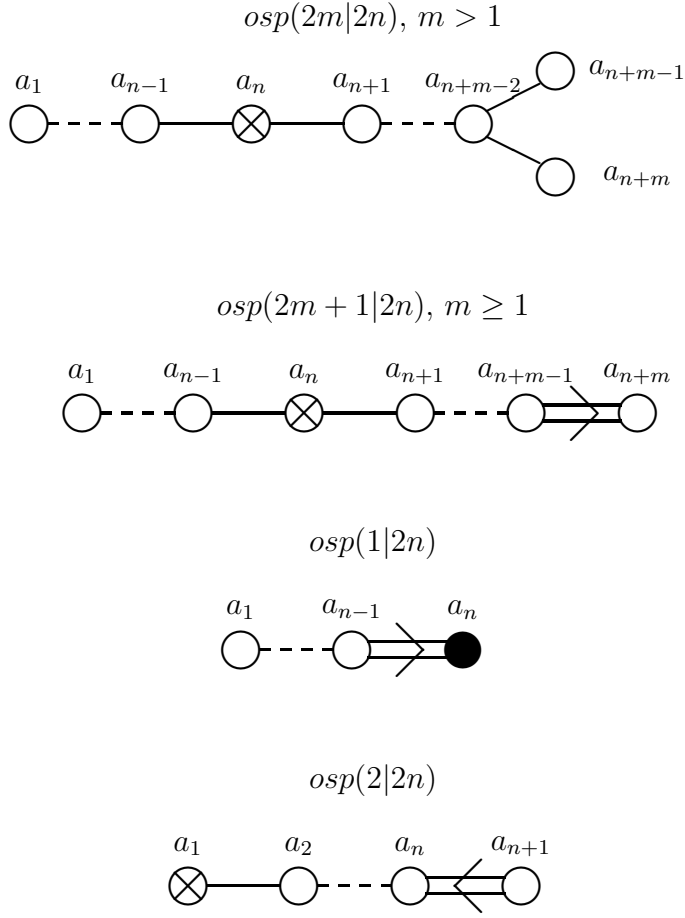

To determine the eigenvalues of the transfer matrix and the
corresponding Bethe Ansatz equations, we use the analytical Bethe
Ansatz method \cite{reshe,mnanal,yabon}. We follow the same
procedure as in \cite{yabon}, by imposing certain constraints on
the eigenvalues, deduced from the crossing symmetry of the model,
the symmetry of the transfer matrix, the analyticity of the
eigenvalues, and the fusion procedure for open spin chains. These
constraints allow to determine the eigenvalues by solving a set of
coupled non-linear consistency equations or Bethe Ansatz
equations.
\\[5mm]
We first describe the case with trivial boundaries,
$K^{-}(\lambda) =K^{+}(\lambda)=1$.
\\
We recall that the fusion procedure for the open spin chain
\cite{doikou,mnfusion} yields the fused transfer matrix
\begin{eqnarray}
  \tilde t(\lambda) = \zeta(2\lambda+2i\kappa)\ t(\lambda)\
  t(\lambda + i\kappa) - \zeta(\lambda+i\kappa)^{2N} q(2\lambda
  +i\kappa)q(-2 \lambda - 3i\kappa)\,, \label{fusion}
\end{eqnarray}
where we define
\begin{eqnarray}
  \zeta(\lambda) =
  (\lambda+i\kappa)(\lambda+i)(\lambda-i\kappa)(\lambda-i)\,, ~~
  q(\lambda) = \theta_0(\lambda - i\theta_0)(\lambda - i\kappa).
\end{eqnarray}
Note that the value of $q(\lambda)$ is related to the specific
choice of the position of the orthogonal and the symplectic part
in the $R$-matrix. We choose for the general case (apart from the
$osp(2|2n)$ case) the symplectic part to be ``outside'' and the
orthogonal part to be ``inside''. This formulation corresponds to
a specific Dynkin diagram: the so-called distinguished one (see
fig. \ref{fig:DD1}). For the case of $osp(2|2n)$ the distinguished
Dynkin diagram has a special form (see fig. \ref{fig:DD1}): it
corresponds to the orthogonal part being ``outside'' and the
symplectic part being inside. These considerations justify the
conventions we have adopted in
(\ref{eq:deftheta})-(\ref{eq:defigrad}) and
(\ref{eq:deftheta2})-(\ref{eq:defigrad2}).\\
From the crossing symmetry of the $R$-matrix (\ref{cu}) it follows
that: $t(\lambda) =t(-\lambda -i\kappa)$. The transfer matrix with
$K^{-}=K^{+} =1$ is obviously $osp(M|2n)$ invariant, since the
corresponding $R$-matrix (\ref{r}) is $osp(M|2n)$ invariant,
namely
\begin{equation}
  \Big [R_{12}, ~~U_{1}+U_{2} \Big ] =0,
\end{equation}
where $U$ is any generator of the $osp(M|2n)$ algebra. Finally,
from the assumption of analyticity of the eigenvalues, we require
that no singularity appears in the Bethe eigenvalues. The
aforementioned set of constraints uniquely fix the eigenvalues.

We now choose an appropriate pseudo-vacuum, which is an exact
eigenstate of the transfer matrix; it is the state with all
``spins'' up, i.e.
\begin{eqnarray}
\vert \omega_{+} \rangle = \bigotimes_{i=1}^{N} \vert + \rangle
_{i} ~~\mbox{where} ~~\vert + \rangle = \left (
\begin{array}{c}
1 \\
0 \\
\vdots \\
0 \\
\end{array}
\right)\,\in\,\CC^{M+2n}\,. \label{pseudo}
\end{eqnarray}
Our choice of $\theta_0$ ensures that this state is always
bosonic,
whichever orthosymplectic superalgebra we consider.\\
After some lengthy computation, we determine explicitly the action
of the transfer matrix as $t(\lambda) \vert \omega_{+} \rangle =
\Lambda^{0}(\lambda) \vert \omega_{+} \rangle$, where
$\Lambda^{0}(\lambda)$ is given by the following expression
\begin{eqnarray}
\Lambda^{0}(\lambda) = a(\lambda)^{2N} g_{0}(\lambda)
+b(\lambda)^{2N}\sum_{l=1}^{2n+M-2} (-1)^{[l+1]} g_{l}(\lambda)+
c(\lambda)^{2N} g_{2n+M-1}(\lambda) \label{eigen0}
\end{eqnarray}
with
\begin{eqnarray}
a(\lambda) =(\lambda+i)(\lambda+i\kappa), ~~b(\lambda) =\lambda
(\lambda+i\kappa), ~~c(\lambda) =\lambda (\lambda+i\kappa-i)
\label{numbers}
\end{eqnarray}
The expressions of the functions $g_l(\lambda)$ depend on the case
we consider.
\\[3mm]
{\bf For the generic $osp(M|2n)$ case, $M\neq2$}, they are given
by (with $M=2m$ or $M=2m+1$):
\begin{eqnarray}
g_{l}(\lambda) &=& {\lambda (\lambda+ {i\kappa \over 2} -{i \over
2})(\lambda+i\kappa) \over (\lambda+ {i \kappa \over
2})(\lambda+{i l \over 2})(\lambda+{i (l +1)\over 2})},
~~l=0,\ldots,n-1,
\non\\
g_{l}(\lambda) &=& {\lambda (\lambda+ {i\kappa \over 2} - {i \over
2})(\lambda+i\kappa) \over (\lambda+ {i \kappa \over
2})(\lambda+{i n}-{i l \over 2})(\lambda+ in - {i (l+1)\over 2})},
~~l=n,\ldots,n+m-1
\non\\
g_{n+m}(\lambda) &=&
\frac{\lambda(\lambda+i\kappa)}{(\lambda+i\frac{n-m}{2})
(\lambda+i\frac{n-m+1}{2})}\qquad \mbox{if } M=2m+1
\non\\
g_{l}(\lambda) &=& g_{2n+M-l-1}(-\lambda -i\kappa),
~~l=0,1,...,M+2n \label{g2}
\end{eqnarray}
In this case, we have $\kappa=n+1-\frac{M}{2}$. We also set $k=n+m$.
\\[3mm]
{\bf For the case of $osp(2|2n)$}, due to the different
conventions, one has:
\begin{eqnarray}
g_{0}(\lambda) &=& {(\lambda+ {i\kappa \over 2} +{i \over
2})(\lambda+i\kappa) \over (\lambda+ {i \kappa \over 2})
(\lambda+{i \over 2})},
\non\\
g_{l}(\lambda) &=& {\lambda (\lambda+ {i\kappa \over 2} + {i \over
2})(\lambda+i\kappa) \over (\lambda+ {i \kappa \over
2})(\lambda+i-{i l \over 2})(\lambda+ i -{i (l +1)\over 2})},
~~l=1,\ldots,n,
\non\\
g_{l}(\lambda) &=& g_{2n+1-l}(-\lambda -i\kappa), ~~l=
0,\ldots,2n+1 \label{g22}
\end{eqnarray}
We remind that in this latter case, $\kappa=-n$.

\subsection{Dressing functions}
{From} the exact expression for the pseudo-vacuum eigenvalue, we
introduce the following assumption for the structure of the
general eigenvalues:
\begin{eqnarray}
\Lambda(\lambda) = a(\lambda)^{2N} g_{0}(\lambda)A_{0}(\lambda)
+b(\lambda)^{2N}\sum_{l=1}^{2n+M-2}
(-1)^{[l+1]}g_{l}(\lambda)A_{l}(\lambda)+ c(\lambda)^{2N}
g_{2n+M-1}(\lambda)A_{2n+M-1}(\lambda)\nonumber\\
\label{eigen}
\end{eqnarray}
where the so-called ``dressing functions'' $A_{i}(\lambda)$ need
now to
be determined. \\
We immediately get from the crossing symmetry of the transfer
matrix:
\begin{eqnarray}
A_{l}(\lambda) =
A_{2n+M-l-1}(-\lambda -i\kappa) \qquad l=0,...,M+2n-1
\,. \label{11}
\end{eqnarray}
Moreover, we obtain from the fusion relation (\ref{fusion}) the
following identity, by a comparison of the forms (\ref{eigen}) for
the initial and fused auxiliary spaces:
\begin{eqnarray}
A_{0}(\lambda+i\kappa)A_{2n+M-1}(\lambda) =1 \,. \label{22}
\end{eqnarray}
Gathering the above two equations (\ref{11}), (\ref{22}) we
conclude
\begin{eqnarray}
A_{0}(\lambda)A_{0}(-\lambda) = 1 \,. \label{33}
\end{eqnarray}
Additional constraints are then imposed on the ``dressing
functions'' from analyticity properties. Studying carefully the
common poles of successive $g_{l}$'s, we deduce from the form of
the $g_{l}$ functions (\ref{g2}) that $g_{l}$ and $g_{l-1}$ have
common poles at $\lambda =-{il\over 2}$ or $\lambda =-in+{il\over
2}$, therefore from analyticity requirements
\begin{eqnarray}
A_{l}(-{il \over 2})&=& A_{l-1}(-{il \over 2}),
~~~l=1,\ldots,n-1, \non\\
A_{l}(-in+{il \over 2})&=& A_{l-1}(-in+{il \over 2}),
~~l=n,\ldots,n+m-1 \label{anal1}
\end{eqnarray}
There is an extra constraint when $M=2m+1$, namely
\begin{eqnarray}
A_{n+m}(-in+{ik \over 2})=A_{n+m-1}(-in+{ik \over 2}) \,.
\label{anal2}
\end{eqnarray}
Having deduced the necessary constraints for the ``dressing
functions'', we determine them explicitly. The ``dressing
functions'' $A_{l}$ are basically characterised by a set of
parameters $\lambda_{j}^{(l)}$ with $j=1, \ldots, M^{(l)}$, where
the integer numbers $M^{(l)}$ are related to the diagonal
generators of $osp(M|2n)$. These generators are defined as:
\begin{eqnarray}
S^{(l)}= \sum_{i=1}^{N} s_{i}^{(l)}, ~~s^{(l)} =(e_{ll}-e_{\bar l \bar l})/2,
~~(e_{kl})_{ij}=\delta_{ik} \delta_{jl}. \label{qn}
\end{eqnarray}
The precise identification of $M^{(l)}$ follows from the symmetry
of the transfer matrix (see also \cite{reshe}):
\begin{eqnarray}
S^{(l)} &=& M^{(l-1)} - M^{(l)}, ~~l=1,\ldots,n-1,n+1,...,n+m-2 \\
S^{(n)} &=& M^{(n-1)} - 2M^{(n)} \\
S^{(n+m-1)} &=& M^{(n+m-2)} - M^{(n+m-1)}, ~~S^{(n+m)} =
M^{(n+m-1)} -
M^{(n+m)}, ~~\mbox{if $M=2m+1$}\qquad\\
S^{(n+m-1)} &=& M^{(n+m-2)} - M^{(+)}- M^{(-)}, ~~S^{(n+m)} =
M^{(+)} - M^{(-)}, ~~\mbox{if $M=2m$ }\quad \label{quant}
\end{eqnarray}
and $M^{(0)} =\frac{N}2$.

\subsection*{A. $\bf osp(2m+1|2n)$}
The dressing functions take the form:
\begin{eqnarray}
A_{0}(\lambda) &=& \prod_{j=1}^{M^{(1)}}{\lambda+
\lambda_{j}^{(1)}-{i\over 2}\over \lambda+ \lambda_{j}^{(1)}
+{i\over 2}}\ {\lambda-\lambda_{j}^{(1)}-{i\over 2} \over
\lambda-\lambda_{j}^{(1)} +{i\over 2}} \,,\non\\
A_{l}(\lambda) &=& \prod_{j=1}^{M^{(l)}}
{\lambda+\lambda_{j}^{(l)}+{il\over 2}+i \over \lambda+
\lambda_{j}^{(l)} +{il\over2}} \;
{\lambda-\lambda_{j}^{(l)}+{il\over 2}+i\over \lambda-
\lambda_{j}^{(l)} +{il\over 2}} \non\\
& \times & \prod_{j=1}^{M^{(l+1)}}{\lambda+
\lambda_{j}^{(l+1)}+{il\over 2}-{i\over 2}\over \lambda+
\lambda_{j}^{(l+1)} +{il\over 2} +{i\over 2}}\
{\lambda-\lambda_{j}^{(l+1)}+{il \over 2}-{i\over 2} \over
\lambda-\lambda_{j}^{(l+1)} + {il\over 2}+{i\over 2}} \,, \qquad l
=
1,\ldots , n-1 \non
\end{eqnarray}
\begin{eqnarray}
A_{l}(\lambda) &=& \prod_{j=1}^{M^{(l)}}
{\lambda+\lambda_{j}^{(l)}+in-{il\over 2}-i \over \lambda+
\lambda_{j}^{(l)} +in-{il\over2}} \;
{\lambda-\lambda_{j}^{(l)}+in-{il\over 2}-i\over \lambda-
\lambda_{j}^{(l)} +in-{il\over 2}} \non\\
& \times & \prod_{j=1}^{M^{(l+1)}}{\lambda+
\lambda_{j}^{(l+1)}+in-{il\over 2}+{i\over 2}\over \lambda+
\lambda_{j}^{(l+1)} +in-{il\over 2} -{i\over 2}}\
{\lambda-\lambda_{j}^{(l+1)}+in-{il \over 2}+{i\over 2} \over
\lambda-\lambda_{j}^{(l+1)} +in- {il\over 2}-{i\over 2}} \,,
\qquad l =
n,\ldots,n+m-1 \non\\
A_{n+m}(\lambda) &=& \prod_{j=1}^{M^{(k)}}
{\lambda+\lambda_{j}^{(k)}+in-{ik\over 2}+i\over \lambda+
\lambda_{j}^{(k)} +in-{ik\over2}} \;
{\lambda-\lambda_{j}^{(k)}+in-{ik\over 2}+i\over \lambda-
\lambda_{j}^{(k)} +in-{ik\over 2}} \non\\ & \times
&{\lambda+\lambda_{j}^{(k)}+in-{ik\over 2}-{i\over 2} \over
\lambda+ \lambda_{j}^{(k)} +in-{ik\over 2}+{i\over 2}} \;
{\lambda-\lambda_{j}^{(k)}+in-{ik\over 2}-{i\over 2}\over \lambda-
\lambda_{j}^{(k)} +in-{ik\over 2}+{i\over 2}}\,, \label{a2}
\end{eqnarray}
and $A_{l}(\lambda) = A_{2n+2m-l}(-\lambda -i\kappa)$ for $l >
n+m$, $\kappa = n-m+{1 \over 2}$.

\subsection*{B. $\bf osp(2m|2n)$ with $\bf m>1$}
The dressing functions are the same as in the previous case for
$l=0, \ldots, n+m-3$, but
\begin{eqnarray}
A_{n+m-2}(\lambda) &=& \prod_{j=1}^{M^{(k-2)}}
{\lambda+\lambda_{j}^{(k-2)}+in-{ik\over 2} \over \lambda+
\lambda_{j}^{(k-2)} +in-{ik\over2}+i} \;
{\lambda-\lambda_{j}^{(k-2)}+in-{ik\over 2} \over \lambda-
\lambda_{j}^{(k-2)} +in-{ik\over 2}+i} \non\\
& \times & \prod_{j=1}^{M^{(+)}}{\lambda+\lambda_{j}^{(+)}+in
-{ik\over 2}+{3i\over 2}\over \lambda+ \lambda_{j}^{(+)}
+in-{ik\over 2}+{i\over 2}}\;
{\lambda-\lambda_{j}^{(+)}+in-{ik\over 2}+{3i\over 2}\over
\lambda-
\lambda_{j}^{(+)} +in-{ik\over 2}+{i\over 2}} \non\\
& \times & \prod_{j=1}^{M^{(-)}}{\lambda+\lambda_{j}^{(-)}+in
-{ik\over 2}+{3i\over 2}\over \lambda+ \lambda_{j}^{(-)}
+in-{ik\over 2}+{i\over 2}} \;
{\lambda-\lambda_{j}^{(-)}+in-{ik\over 2}+{3i\over 2}\over
\lambda-
\lambda_{j}^{(-)} +in-{ik\over 2}+{i\over 2}}\,, ~~ \non\\
A_{n+m-1}(\lambda) &=& \prod_{j=1}^{M^{(+)}}
{\lambda+\lambda_{j}^{(+)}+in-{ik\over 2}+{3i\over 2}\over
\lambda+ \lambda_{j}^{(+)} +in-{ik\over2}+{i \over 2}} \;
{\lambda-\lambda_{j}^{(+)}+in-{ik\over 2} +{3i \over 2}\over
\lambda-
\lambda_{j}^{(+)} +in-{ik\over 2}+{i \over 2}} \non\\
& \times & \prod_{j=1}^{M^{(-)}}
{\lambda+\lambda_{j}^{(-)}+in-{ik\over 2}-{i\over 2}\over \lambda+
\lambda_{j}^{(-)} +in- {ik\over 2}+{i\over 2}} \;
{\lambda-\lambda_{j}^{(-)}+in-{ik\over 2}-{i\over 2}\over \lambda-
\lambda_{j}^{(-)} +in-{ik\over 2}+{i\over 2}}\, \label{a3}
\end{eqnarray}
and $A_{l}(\lambda) = A_{2n+2m-l-1}(-\lambda -i\kappa)$ for $l
>n+m-1$, $\kappa=n+1-m$.

\subsection*{C. $\bf osp(2|2n)$}
As already mentioned, this case must be treated separately,
because of the different position of the orthogonal and symplectic
parts in the $R$-matrix. Here the orthogonal part of the
$R$-matrix is considered to be ``outside'' and the symplectic part
``inside'' as opposed to the previous cases. The corresponding
dressing functions have the form
\begin{eqnarray}
A_{0}(\lambda) &=& \prod_{j=1}^{M^{(1)}}{\lambda+
\lambda_{j}^{(1)}-{i\over 2}\over \lambda+ \lambda_{j}^{(1)}
+{i\over 2}}\ {\lambda-\lambda_{j}^{(1)}-{i\over 2} \over
\lambda-\lambda_{j}^{(1)} +{i\over 2}} \,, \non\\
A_{l}(\lambda) &=& \prod_{j=1}^{M^{(l)}}
{\lambda+\lambda_{j}^{(l)}-{il\over 2} \over \lambda+
\lambda_{j}^{(l)} +i-{il\over2}} \;
{\lambda-\lambda_{j}^{(l)}-{il\over 2}\over \lambda-
\lambda_{j}^{(l)} +i-{il\over 2}} \non\\
& \times & \prod_{j=1}^{M^{(l+1)}}{\lambda+
\lambda_{j}^{(l+1)}+{3i\over 2}-{il\over 2}\over \lambda+
\lambda_{j}^{(l+1)} +{i\over 2}-{il\over 2}}\
{\lambda-\lambda_{j}^{(l+1)}+{3i\over 2}-{il \over 2} \over
\lambda-\lambda_{j}^{(l+1)} +{i\over 2}- {il\over 2}} \,, \qquad
l= 1,\ldots,n-1 \non\\ A_{n}(\lambda) &=& \prod_{j=1}^{M^{(n)}}
{\lambda+\lambda_{j}^{(n)}-{in\over 2}\over
\lambda+\lambda_{j}^{(n)} +i- {in\over2}} \;
{\lambda-\lambda_{j}^{(n)}-{in\over 2}\over \lambda-
\lambda_{j}^{(n)} +i-{in\over 2}} \non\\ & \times &
\prod_{j=1}^{M^{(n+1)}}{\lambda+\lambda_{j}^{(n+1)}+2i-{in\over 2}
\over \lambda+ \lambda_{j}^{(n+1)} -{in\over 2}} \;
{\lambda-\lambda_{j}^{(n+1)}+2i-{in\over 2}\over \lambda-
\lambda_{j}^{(n+1)} -{in\over 2}}\,, \label{a4}
\end{eqnarray}
and $A_{l}(\lambda) = A_{2n+1-l}(-\lambda -i\kappa)$ for $l>n$,
$\kappa=-n$.

\subsection{Bethe Ansatz equations}
We define the function
\begin{equation}
  \label{eq:ex}
  e_x(\lambda) = \frac{\lambda +\frac{ix}{2}}{\lambda - \frac{ix}{2}} \;.
\end{equation}
{}From the analyticity requirements one obtains the Bethe Ansatz
equations which read as:

\subsection*{A. $\bf osp(2m+1|2n)$}
\begin{eqnarray}
e_{1}(\lambda_{i}^{(1)})^{2N} &\!\!=\!\!& \prod_{j=1,j \ne
i}^{M^{(1)}} e_{2}(\lambda_{i}^{(1)} - \lambda_{j}^{(1)})\
e_{2}(\lambda_{i}^{(1)} + \lambda_{j}^{(1)})\ \prod_{
j=1}^{M^{(2)}}e_{-1}(\lambda_{i}^{(1)} -
\lambda_{j}^{(2)})\ e_{-1}(\lambda_{i}^{(1)} + \lambda_{j}^{(2)})\,, \non\\
1 &\!\!=\!\!& \prod_{j=1,j \ne i}^{M^{(l)}}
e_{2}(\lambda_{i}^{(l)} - \lambda_{j}^{(l)})\
e_{2}(\lambda_{i}^{(l)} + \lambda_{j}^{(l)})\ \prod_{ \tau = \pm
1}\prod_{ j=1}^{M^{(l+\tau)}}e_{-1}(\lambda_{i}^{(l)} -
\lambda_{j}^{(l+\tau)})\ e_{-1}(\lambda_{i}^{(l)} +
\lambda_{j}^{(l+\tau)}) \non\\
&& l= 2,\ldots,n+m-1, \;\; l \neq n \non\\
1 &\!\!=\!\!& \prod_{j=1}^{M^{(n+1)}} e_{1}(\lambda_{i}^{(n)} -
\lambda_{j}^{(n+1)})\ e_{1}(\lambda_{i}^{(n)} +
\lambda_{j}^{(n+1)})\ \prod_{
j=1}^{M^{(n-1)}}e_{-1}(\lambda_{i}^{(n)} - \lambda_{j}^{(n-1)})\
e_{-1}(\lambda_{i}^{(n)} + \lambda_{j}^{(n-1)}) \non\\
1 &\!\!=\!\!& \prod_{j=1,j \ne i}^{M^{(n+m)}}
e_{1}(\lambda_{i}^{(n+m)} - \lambda_{j}^{(n+m)})\
e_{1}(\lambda_{i}^{(n+m)} + \lambda_{j}^{(n+m)})\
\non\\
& \times &\prod_{j=1}^{M^{(n+m-1)}}e_{-1}(\lambda_{i}^{(n+m)} -
\lambda_{j}^{(n+m-1)})\ e_{-1}(\lambda_{i}^{(n+m)} +
\lambda_{j}^{(n+m-1)}) \label{BAE1}
\end{eqnarray}
In particular for $M=1$ the Bethe Ansatz equations become
\begin{eqnarray}
e_{1}(\lambda_{i}^{(1)})^{2N} &\!\!=\!\!& \prod_{j=1,j \ne
i}^{M^{(1)}} e_{2}(\lambda_{i}^{(1)} - \lambda_{j}^{(1)})\
e_{2}(\lambda_{i}^{(1)} + \lambda_{j}^{(1)})\ \prod_{
j=1}^{M^{(2)}}e_{-1}(\lambda_{i}^{(1)} -
\lambda_{j}^{(2)})\ e_{-1}(\lambda_{i}^{(1)} + \lambda_{j}^{(2)})\,, \non\\
1 &\!\!=\!\!& \prod_{j=1,j \ne i}^{M^{(l)}}
e_{2}(\lambda_{i}^{(l)} - \lambda_{j}^{(l)})\
e_{2}(\lambda_{i}^{(l)} + \lambda_{j}^{(l)})\ \prod_{ \tau = \pm
1}\prod_{ j=1}^{M^{(l+\tau)}}e_{-1}(\lambda_{i}^{(l)} -
\lambda_{j}^{(l+\tau)})\ e_{-1}(\lambda_{i}^{(l)} +
\lambda_{j}^{(l+\tau)}) \non\\ &&l= 2,\ldots,n-1,
\non\\
1 &\!\!=\!\!& \prod_{j=1,j \ne i}^{M^{(n)}}
e_{-1}(\lambda_{i}^{(n)} - \lambda_{j}^{(n)})\
e_{-1}(\lambda_{i}^{(n)} + \lambda_{j}^{(n)})
e_{2}(\lambda_{i}^{(n)} - \lambda_{j}^{(n)})\
e_{2}(\lambda_{i}^{(n)} + \lambda_{j}^{(n)})\non\\
& \times & \prod_{ j=1}^{M^{(n-1)}}e_{-1}(\lambda_{i}^{(n)} -
\lambda_{j}^{(n-1)})\ e_{-1}(\lambda_{i}^{(n)} +
\lambda_{j}^{(n-1)}) \label{BAE1b}
\end{eqnarray}

\subsection*{B. $\bf osp(2m|2n)$ with $\bf m>1$}
The first $n+m-3$ equations are the same as in the previous case
for $M=2m+1$, see eq. (\ref{BAE1}), but the last three equations
are modified, and they become identical to the last three
equations of the $so(2n+2m)$ open spin chain, namely,
\begin{eqnarray}
1 &\!\!=\!\!& \prod_{j=1,j \ne i}^{M^{(n+m-2)}}
e_{2}(\lambda_{i}^{(n+m-2)} - \lambda_{j}^{(n+m-2)})\
e_{2}(\lambda_{i}^{(n+m-2)} + \lambda_{j}^{(n+m-2)})\ \non\\
& \times & \prod_{ j=1}^{M^{(n+m-3)}}e_{-1}(\lambda_{i}^{(n+m-2)}
- \lambda_{j}^{(n+m-3)})\ e_{-1}(\lambda_{i}^{(n+m-2)} +
\lambda_{j}^{(n+m-3)})\non\\
&\times & \prod_{ \tau =\pm}\prod_{
j=1}^{M^{(\tau)}}e_{-1}(\lambda_{i}^{(n+m-2)} -
\lambda_{j}^{(\tau)})\
e_{-1}(\lambda_{i}^{(n+m-2)} + \lambda_{j}^{(\tau)}) \non\\
1 &\!\!=\!\!& \prod_{j=1,j \ne i}^{M^{(\tau)}}
e_{2}(\lambda_{i}^{(\tau)} - \lambda_{j}^{(\tau)})\
e_{2}(\lambda_{i}^{(\tau)} + \lambda_{j}^{(\tau)})\ \prod_{
j=1}^{M^{(n+m-2)}}e_{-1}(\lambda_{i}^{(\tau)} -
\lambda_{j}^{(n+m-2)})\ e_{-1}(\lambda_{i}^{(\tau)} +
\lambda_{j}^{(n+m-2)}) \non\\
\label{BAE2}
\end{eqnarray}

\subsection*{C. $\bf osp(2|2n)$}
\begin{eqnarray}
e_{1}(\lambda_{i}^{(1)})^{2N} &\!\!=\!\!& \ \prod_{
j=1}^{M^{(2)}}e_{1}(\lambda_{i}^{(1)} -
\lambda_{j}^{(2)})\ e_{1}(\lambda_{i}^{(1)} + \lambda_{j}^{(2)})\,, \non\\
1 &\!\!=\!\!& \prod_{j=1,j \ne i}^{M^{(l)}}
e_{2}(\lambda_{i}^{(l)} - \lambda_{j}^{(l)})\
e_{2}(\lambda_{i}^{(l)} + \lambda_{j}^{(l)})\ \prod_{ \tau = \pm
1}\prod_{ j=1}^{M^{(l+\tau)}}e_{-1}(\lambda_{i}^{(l)} -
\lambda_{j}^{(l+\tau)})\ e_{-1}(\lambda_{i}^{(l)} +
\lambda_{j}^{(l+\tau)}) \non\\ &&l= 2, \ldots ,n-1,
\non
\end{eqnarray}
\begin{eqnarray}
1 &\!\!=\!\!& \prod_{j=1,}^{M^{(n+1)}} e_{-2}(\lambda_{i}^{(n)} -
\lambda_{j}^{(n+1)})\
e_{-2}(\lambda_{i}^{(n)} + \lambda_{j}^{(n+1)})\ \non\\
& \times & \prod_{ j=1}^{M^{(n-1)}}e_{-1}(\lambda_{i}^{(n)} -
\lambda_{j}^{(n-1)})\ e_{-1}(\lambda_{i}^{(n)} +
\lambda_{j}^{(n-1)})\ \prod_{j=1,j \ne i}^{M^{(n)}}
e_{2}(\lambda_{i}^{(n)} -\lambda_{j}^{(n)})\
e_{2}(\lambda_{i}^{(n)} + \lambda_{j}^{(n)}) \non\\
1 &\!\!=\!\!& \prod_{j=1,j \ne i}^{M^{(n+1)}}
e_{4}(\lambda_{i}^{(n+1)} - \lambda_{j}^{(n+1)})\
e_{4}(\lambda_{i}^{(n+1)} + \lambda_{j}^{(n+1)})\ \non\\ &&
\prod_{ j=1}^{M^{(n)}}e_{-2}(\lambda_{i}^{(n+1)} -
\lambda_{j}^{(n)})\ e_{-2}(\lambda_{i}^{(n+1)} +
\lambda_{j}^{(n)}). \label{BAE3}
\end{eqnarray}
In particular the equations for the $osp(2|2)$ open chain are
given by
\begin{eqnarray}
e_{1}(\lambda_{i}^{(1)})^{2N} &\!\!=\!\!& \ \prod_{
j=1}^{M^{(2)}}e_{2}(\lambda_{i}^{(1)} -
\lambda_{j}^{(2)})\ e_{2}(\lambda_{i}^{(1)} + \lambda_{j}^{(2)})\,, \non\\
1 &\!\!=\!\!& \prod_{j=1}^{M^{(1)}} e_{-2}(\lambda_{i}^{(2)} -
\lambda_{j}^{(1)})\ e_{-2}(\lambda_{i}^{(2)} + \lambda_{j}^{(1)})\
\prod_{ j=1,j \ne i}^{M^{(2)}}e_{4}(\lambda_{i}^{(2)} -
\lambda_{j}^{(2)})\ e_{4}(\lambda_{i}^{(2)} + \lambda_{j}^{(2)})
\label{osp2}
\end{eqnarray}
Notice that there is a one-to-one correspondence between the
distinguished Dynkin diagrams (see fig. 1) and the Bethe Ansatz
equations derived for each case. The Bethe Ansatz equations for
$osp(1|2n)$, $osp(2|2n)$, $osp(2m|2)$, $osp(2m+1|2)$ can now be
compared with the corresponding results obtained in \cite{mara}
for super spin chains with periodic boundaries. Let us point out
however that we derived explicitly the Bethe Ansatz equations for
any $osp(M|2n)$ open spin chain, and we expect the corresponding
equations for a chain with periodic boundary conditions to be
``halved'' (i.e. half of the factors in the products should be
missing) compared to the ones we found.

\subsection{Non-trivial diagonal boundary conditions}
Until now we have derived the Bethe Ansatz equations for trivial
boundary conditions, namely $K^{-}=K^{+}=1$. We shall now insert
non-trivial boundary effects and then rederive the modified Bethe
Ansatz equations. We choose $K^{-}$ to be one of the diagonal
solutions D1, D2, D3 found in Proposition 3.1 of \cite{yabon}. We
consider, for simplicity but without loss of generality, $K^{+}
=1$. Note that the pseudo-vacuum remains an exact eigenstate after
this modification. We rewrite the solutions D1, D2 and D3 of
\cite{yabon} in a slightly modified notation, which we are going
to use from now on.

\medskip

{\bf D1:} The solution D1 can be written in the following form
\begin{eqnarray}
K(\lambda) = diag( \alpha, \ldots ,\alpha, \beta, \dots, \beta)
\,. \label{eq:solD1}
\end{eqnarray}
The number of $\alpha 's$ is equal to the number of $\beta 's$, so
that this solution exists only for the $osp(2m|2n)$ cases as
stated in Proposition 3.1 of ref. \cite{yabon}, and
\begin{eqnarray}
\alpha(\lambda) = -\lambda +i\xi, ~~\beta(\lambda) = \lambda
+i\xi,
\end{eqnarray}
where $\xi$ is the free boundary parameter.

\medskip

{\bf D2:} Solution D2 can be written in the $osp(M|2n)$ case
($M>2$) as
\begin{eqnarray}
K(\lambda) = diag( \underbrace{1, \ldots, 1}_{n}, \bar\alpha,
\underbrace{1, \ldots, 1}_{M-2}, \bar\beta, \underbrace{1, \ldots,
1}_{n} ) \label{eq:solD2}
\end{eqnarray}
and for the $osp(2|2n)$ case as
\begin{eqnarray}
K(\lambda) = diag( \bar\alpha, \underbrace{1, \ldots, 1}_{2n},
\bar\beta ) \label{eq:solD2bis}
\end{eqnarray}
with
\begin{eqnarray}
\bar\alpha(\lambda) = { -\lambda +i\xi_{1} \over \lambda
+i\xi_{1}}, ~~\bar\beta(\lambda) = { -\lambda +i\xi_{n} \over
\lambda +i\xi_{n}} \,,
\end{eqnarray}
where $\xi_{1}$ and $\xi_{n}$ are the boundary parameters which
satisfy the constraint
\begin{eqnarray}
\xi_{1} +\xi_{n} = \kappa - \theta_{0}.
\end{eqnarray}
Obviously, this solution does not exist for the $osp(1|2n)$
superalgebras.

\medskip

{\bf D3:} Solution D3 has the form in the $osp(M|2n)$ case ($M \ne
2$)
\begin{eqnarray}
K(\lambda) = diag( \underbrace{\beta, \dots, \beta}_{n-n_{1}},
\underbrace{\alpha, \ldots ,\alpha}_{n_{1}+m_{1}},
\underbrace{\beta, \dots, \beta}_{M-2m_{1}}, \underbrace{\alpha,
\ldots ,\alpha}_{n_{1}+m_{1}}, \underbrace{\beta, \dots,
\beta}_{n-n_{1}} ). \label{eq:solD3bis}
\end{eqnarray}
while for the $osp(2|2n)$ case it takes the form
\begin{eqnarray}
K(\lambda) = diag( \alpha, \underbrace{\alpha, \ldots
,\alpha}_{n_{1}}, \underbrace{\beta, \dots, \beta}_{2n-2n_{1}},
\underbrace{\alpha, \ldots ,\alpha}_{n_{1}}, \alpha). \label{eq:3}
\end{eqnarray}
The $osp(1|2n)$ case is recovered by taking $M=1$ and $m_{1}=0$ in
(\ref{eq:solD3bis}). \\
Again, $\alpha$ and $\beta$ are given by
\begin{eqnarray}
\alpha(\lambda) = -\lambda +i\xi, ~~\beta(\lambda) = \lambda +i\xi
\end{eqnarray}
where $\xi = (\kappa+2m_{1}-2n_{1}-1)/2$ has a fixed value, the
integers $m_{1}$ and $n_{1}$ being restricted to $0 \le n_{1} \le
n$ and $0 \le m_{1} \le m$ ($M=2m$ or $M=2m+1$) respectively,
$n_{1}$ and $m_{1}$ being neither both zero nor taking maximal
values simultaneously.

\null

We now come to the explicit expression of the eigenvalues when
$K^-$ is one of the above mentioned solutions. We should point out
that the dressing functions are related to the bulk behaviour of
the chain and thus they are form-invariant under changes of
boundary conditions. Indeed the only modifications in the
expression of the eigenvalues (\ref{eigen}) occur in the $g_{l}$
functions, which basically characterise the boundary effects. We
call the new $g_{l}$ functions $\tilde g_{l}$.

\medskip

{\bf D1:} As already mentioned, the solution D1 can only be
applied to $osp(2m|2n)$ with $m \ge 1$. In this case we have
\begin{eqnarray}
\tilde g_{l}(\lambda) &=& (-\lambda + i\xi)g_{l}(\lambda), ~~l=0,
\ldots
,n+m-1 \nonumber \\
\tilde g_{l}(\lambda) &=& (\lambda + i\xi +i\kappa)g_{l}(\lambda),
~~l=n+m, \ldots, 2n+2m-1 \label{eq:tg1}
\end{eqnarray}
where $g_l(\lambda)$ are given by (\ref{g2})--(\ref{g22}). The
system with such boundaries has a residual symmetry 
$sl(m|n)$, which immediately follows from the structure
of the corresponding $K$ matrix.

\medskip

{\bf D2:} We have to separate the cases $osp(M|2n)$ with $M \ne 2$
and $osp(2|2n)$. In the $osp(M|2n)$ case with $M \ne 2$, one gets
\begin{eqnarray}
\tilde g_{l}(\lambda) &=& g_{l}(\lambda), ~~l=0, \ldots, n-1,
\qquad \tilde g_{n}(\lambda) \;\;=\;\; {(-\lambda + i\xi_{1} - in)
\over (\lambda +
i\xi_{1})} \; g_{n}(\lambda), \nonumber \\
\tilde g_{l}(\lambda) &=& {(\lambda + i\xi_{1} - i) \over (\lambda
+ i\xi_{1})} \; g_{l}(\lambda),
~~l=n+1, \ldots, n+M-2 \nonumber \\
\tilde g_{n+M-1}(\lambda) &=& {(\lambda + i\xi_{1} - i) \over
(\lambda + i\xi_{1})} \; {(-\lambda + i\xi_{n} - in + iM - 3i)
\over (\lambda +
i\xi_{n})} \; g_{n+M-1}(\lambda), \nonumber \\
\tilde g_{l}(\lambda) &=& {(\lambda + i\xi_{1} - i) \over (\lambda
+ i\xi_{1})} \; {(\lambda + i\xi_{n} - i) \over (\lambda +
i\xi_{n})} \; g_{l}(\lambda), ~~l=n+M, \ldots, 2n+M-1
\label{eq:tg2}
\end{eqnarray}
In the $osp(2|2n)$ case, the formulae are similar to the $sp(2n)$
case:
\begin{eqnarray}
\tilde g_{0}(\lambda) &=& {(-\lambda + i\xi_{1}) \over (\lambda +
i\xi_{1})} \; g_{0}(\lambda), \qquad \tilde g_{2n+1}(\lambda)
\;\;=\;\; {(\lambda + i\xi_{1} +i)\over (\lambda +
i\xi_{1})}{(\lambda + i\xi_{1}+i\kappa) \over (-\lambda -i\kappa+
i\xi_{1}+i)} \;
g_{2n+1}(\lambda), \nonumber \\
\tilde g_{l}(\lambda) &=& {(\lambda + i\xi_{1}+i) \over (\lambda +
i\xi_{1})} \; g_{l}(\lambda), ~~l=1, \ldots, 2n \label{eq:tg22}
\end{eqnarray}
The functions $g_{l}(\lambda)$ are given by
(\ref{g2})--(\ref{g22}).
The system with such boundaries has a residual symmetry 
$osp(M-2|2n)$.

\medskip

{\bf D3:} For the D3 solution we find the following modified $g$
functions in the case of $osp(M|2n)$ with $M \ne 2$:
\begin{eqnarray}
\tilde g_{l}(\lambda) &=& (\lambda +i\xi) \; g_{l}(\lambda),
~~l=0,\ldots,
n-n_{1}-1 \nonumber \\
\tilde g_{l}(\lambda) &=& (-\lambda +i\xi -in +in_{1}) \;
g_{l}(\lambda),
~~l=n-n_{1}, \ldots, n+m_{1}-1 \nonumber \\
\tilde g_{l}(\lambda) &=& (\lambda +i{\kappa \over 2} -{i \over
2}) \; g_{l}(\lambda),
~~l=n+m_{1}, \ldots, n+m-1 \nonumber \\
\tilde g_{M+2n-1-l}(\lambda) &=& {(\lambda +i\frac{\kappa}{2}
-\frac{i}{2}) \over (\lambda +i\frac{\kappa}{2} +\frac{i}{2}) }
\; \tilde g_{l}(-\lambda -i\kappa), ~~l=0, \ldots, n+m-1 \nonumber \\
\label{eq:tg3}
\end{eqnarray}
and in the case of $osp(2m+1|2n)$
\begin{eqnarray}
\tilde g_{n+m}(\lambda) &=&  (\lambda +i{\kappa \over 2} -{i \over
2}) \; g_{n+m}(\lambda) \label{eq:tg3bis}
\end{eqnarray}
In the $osp(2|2n)$ case, the formulae become
\begin{eqnarray}
\tilde g_{l}(\lambda) &=& (-\lambda +i\xi) \; g_{l}(\lambda),
~~l=0,\ldots,
n_{1} \nonumber \\
\tilde g_{l}(\lambda) &=& (\lambda +i\frac{\kappa}{2} +
\frac{i}{2}) \;
g_{l}(\lambda), ~~l=n_{1}+1,\ldots ,2n-n_{1} \nonumber \\
\tilde g_{l}(\lambda) &=& (-\lambda -i\kappa -i\xi) \;
\frac{(\lambda +i\frac{\kappa}{2} + \frac{i}{2})}{(\lambda
+i\frac{\kappa}{2} - \frac{i}{2})} \; g_{l}(\lambda),
~~l=2n-n_{1}+1,\ldots, 2n+1 \label{eq:tg33}
\end{eqnarray}
The system with such boundaries has a residual symmetry 
$osp(M-2m_1|2n-2n_1) \oplus osp(2m_1|2n_1)$.

\medskip

We now formulate the Bethe Ansatz equations for the general
diagonal solutions. The only modifications induced on Bethe Ansatz
equations are the following for each solution:
\begin{itemize}
\item[{\bf D1}]
The factor $-e_{2\xi+\kappa}^{-1}(\lambda)$ appears in the LHS of
the $(n+m)^{th}$ Bethe equation.
\item[{\bf D2}]
The factor $-e_{2\xi_{1}-n}(\lambda)$ appears in the LHS of the
$n^{th}$
Bethe equation. \\
The factor $-e_{2\xi_{1}-n-1}^{-1}(\lambda)$ appears in the LHS of
the $(n+1)^{th}$ Bethe equation.
\item[{\bf D3}]
The factor $-e_{2\xi-(n-n_{1})}(\lambda)$ appears in the LHS of
the
$(n-n_{1})^{th}$ Bethe equation. \\
The factor $-e_{2\xi+2n_{1}-m_{1}-n}^{-1}(\lambda)$ appears in the
LHS of the $(n+m_{1})^{th}$ Bethe equation.
\end{itemize}

\section{ Scattering for the $osp(1|2n)$ open spin chain}
\setcounter{equation}{0}

\subsection{Low lying excitations}
Before we derive explicitly the bulk and boundary scattering
amplitudes for the $osp(1|2n)$ case we first need to determine the
ground state and the low-lying excitations of the model. We recall
that the energy is derived via the relation $H={d\over d
\lambda}t(\lambda) \vert_ {\lambda = 0}$. It is given by
\begin{eqnarray}
E = -{1\over 2 \pi } \sum_{j=1}^{M^{(1)}} {1\over
{\left(\lambda_{j}^{(1)}\right)}^{2} + {1\over 4}} \,.
\label{energy}
\end{eqnarray}
In what follows we write the Bethe Ansatz equations for the ground
state and the low-lying excitations (holes) of the models under
study. Bethe Ansatz equations may in general only be solved in the
thermodynamic limit $N\to \infty$. In this limit, a state is
described in particular by the density functions
$\sigma^{l}(\lambda)$ of the parameters $\lambda_{i}^{(l)}$.

\subsection*{A. $ \bf osp(1|2)$}
Let us first consider the $osp(1|2)$ case, for which the ground
state consists of one filled Dirac sea with real strings (all
$\lambda_{i}$'s real). The set of Bethe Ansatz equations for the
$osp(1|2)$ case takes the form
\begin{equation}
  e_{1}(\lambda_{i})^{2N}e_{1}(\lambda_{i})e_{-{1\over 2}}(\lambda_{i})
  = \prod_{j=1}^{M} e_{2}(\lambda_{i}-
  \lambda_{j})\ e_{2}(\lambda_{i} + \lambda_{j})\ e_{-1}(\lambda_{i}
  - \lambda_{j})\ e_{-1}(\lambda_{i} + \lambda_{j})\,,\label{baeos}
\end{equation}
The reason why we study this case separately is basically because
we wish to point out the striking similarity between the latter
Bethe Ansatz equations (\ref{baeos}) and the corresponding
equations appearing in the study of the $SU(3)$ open spin chain
with ``soliton non-preserving'' boundary conditions \cite{doikou}.
This is indeed a remarkable connection, which can presumably be
extended to open spin chains with ``soliton non-preserving''
boundary conditions, for higher rank algebras. In particular, we
expect that for any $SU(n)$ ($n$ odd) chain with ``soliton
non-preserving'' boundary conditions the resulting Bethe Ansatz
equations will have the same form as in the $osp(1|n-1)$ open spin
chain with certain diagonal boundaries. We hope to report on this
in detail elsewhere.

We now study the low-lying excitations, which are holes in the
filled Dirac sea. In order to convert the sums into integrals,
after taking the thermodynamic limit ($ N \to \infty $), we employ
the following approximate relation
\begin{eqnarray}
{1\over N} \sum_{i=1}^{M} f(\lambda_{i}^{(l)})= \int_{0}^{\infty}
d\lambda f(\lambda) \sigma^{l}(\lambda) - {1\over N}
\sum_{i=1}^{\nu^{(l)}} f(\tilde \lambda_{i}^{(l)}) -{1\over
2N}f(0) \label{mac}
\end{eqnarray}
where the correction terms take into account the $\nu^{(l)}$ holes
located at values $\tilde \lambda_{i}^{(l)}$ and the halved
contribution at $0^+$. For $osp(1|2)$ in particular $\nu^{(l)}
\equiv \nu$ and $\tilde
\lambda_{i}^{(l)}\equiv \tilde \lambda_{i}$. \\
We shall denote by $\hat{f}(\omega)$ the Fourier transform of any
function $f(\lambda)$. Once we take the logarithm and the
derivative of (\ref{baeos}), we derive the densities from the
equation
\begin{eqnarray}
\hat {\cal K}(\omega) \hat \sigma(\omega) = \hat a_{1}(\omega)
+{1\over N} \hat F(\omega) \label{sigma0}
\end{eqnarray}
where $\displaystyle a_{\ell}(\lambda) = \frac{i}{2\pi} \;
\frac{d}{d\lambda} \, \ln e_{\ell}(\lambda)$ and $\hat
a_{\ell}(\omega) =e^{-{\ell \omega \over 2}}$. Moreover
\begin{equation}
\hat {\cal K}(\omega) = e^{-{\omega \over 2}}{\cosh{3\omega \over
4} \over \cosh{\omega \over 4}},~~F(\lambda)
=a_{2}(\lambda)-a_{{1\over 2}}(\lambda)+\sum_{i=1}^{\nu}
\Big((a_{2}-a_{1})(\lambda
-\tilde\lambda_{i})+(a_{2}-a_{1})(\lambda +\tilde
\lambda_{i})\Big). \label{f1}
\end{equation}
 Finally equation (\ref{sigma0})
can be written as
\begin{eqnarray}
\sigma(\lambda) = 2\epsilon(\lambda) +{1\over N} \Phi(\lambda),
\label{sigma2}
\end{eqnarray}
where
\begin{equation}
\hat \epsilon(\omega) = \hat a_{1}(\omega)\hat {\cal
R}(\omega),~~\hat \Phi(\omega) = \hat {\cal R}(\omega)\hat
F(\omega), ~~\mbox{and} ~~\hat {\cal R}(\omega) = \hat {\cal
K}^{-1}(\omega) \label{ef1}
\end{equation}
 In particular the energy
$\epsilon$ can be written in terms of hyperbolic functions as
$\hat \epsilon(\omega) = {\cosh {\omega \over 4} \over \cosh
{3\omega \over 4}}$.

\subsection*{B. $ \bf osp(1|2n)$}
We recall that we can only consider the D3 solution in this case.
The ground state consists of $n$ filled Dirac seas with real
strings. With the help of relation (\ref{mac}) we derive the
densities that describe the state with $\nu^{(l)}$ holes in the
$l$ sea from the equation
\begin{eqnarray}
\hat {\cal K}(\omega) \hat \sigma(\omega) = \hat a(\omega)
+{1\over N} \hat F(\omega) +{1\over N} \hat G(\omega,\
\xi)\label{sigma}
\end{eqnarray}
where we have introduced
\begin{eqnarray}
a(\lambda) = \left (\begin{array}{c}
2a_{1}(\lambda) \\
0 \\
\vdots \\
0 \\
\end{array}
\right)\,, ~~\sigma(\lambda) = \left (
\begin{array}{c}
\sigma^{1}(\lambda) \\
\vdots \\
\sigma^{l}(\lambda)\\
\vdots \\
\sigma^{n}(\lambda) \\
\end{array}
\right)\,. \label{cols}
\end{eqnarray}
$F(\lambda)$ is a $n$-vector as well with
\begin{eqnarray}
F^{j}(\lambda) &\!\!=\!\!&
a_{1}(\lambda)\delta_{j1}-a_{1}(\lambda) +a_{2}(\lambda)+
\sum_{i=1}^{\nu^{(l)}} \Big(a_{2}(\lambda -\tilde
\lambda_{i}^{(l)})+a_{2}(\lambda +\tilde \lambda_{i}^{(l)})\Big)
\delta_{lj} \non\\
&-&\sum_{i=1}^{\nu^{(l)}} \Big(a_{1}(\lambda -\tilde
\lambda_{i}^{(l)})+a_{1}(\lambda +\tilde \lambda_{i}^{(l)})\Big)
(\delta_{j,l+1} +\delta_{j,l-1}), \qquad (j=1,\ldots ,n-1) \non\\
F^{n}(\lambda) &\!\!=\!\!& a_{1}(\lambda)-a_{{1 \over 2}}(\lambda)
+ a_{2}(\lambda)+ \sum_{i=1}^{\nu^{(l)}} \Big( (a_{2} -
a_{1})(\lambda -\tilde \lambda_{i}^{(l)})+(a_{2}- a_{1})(\lambda
+\tilde \lambda_{i}^{(l)})\Big)
\delta_{l n} \non\\
&-&\sum_{i=1}^{\nu^{(l)}} \Big(a_{1}(\lambda -\tilde
\lambda_{i}^{(l)})+a_{1}(\lambda +\tilde
\lambda_{i}^{(l)})\Big)\delta_{l,n-1} \label{ef2}
\end{eqnarray}
and $\hat {\cal K}$ is a $n\times n$ matrix with entries given by
\begin{eqnarray}
&& \hat {\cal K}_{ij}(\omega) = (1+\hat a_{2}(\omega))\delta_{ij}
- \hat
a_{1}(\omega)(\delta_{i,j+1}+\delta_{i,j-1}), ~~i,j =1,\ldots n-1, \non\\
&& \hat {\cal K}_{n n-1}(\omega) = \hat {\cal K}_{ n-1 n}(\omega)=
\hat a_{1}(\omega), ~~ \hat {\cal K}_{n n}(\omega)= 1-\hat
a_{1}(\omega)+\hat a_{2}(\omega) \label{K2}
\end{eqnarray} Finally, the $n$-vector $G$ carries all the explicit dependence on
the boundary parameter $\xi$ of the D3 solution
\begin{eqnarray}
G^{j}(\lambda) = a_{2\xi -\tilde n_{1}}(\lambda)\ \delta_{j,\tilde
n_{1}}
\end{eqnarray}
where $\tilde n_{1} = n-n_{1}$. Solving equation (\ref{sigma}) we
find the densities $\sigma^{i}$ which describe a Bethe Ansatz
state. The solution of (\ref{sigma}) has the following form
\begin{eqnarray}
\sigma(\lambda) = 2 \epsilon(\lambda) +{1\over N}
\Phi_{0}(\lambda)+{1\over N} \Phi_{1}(\lambda,\ \xi)
\label{sigma3}
\end{eqnarray}
where $\epsilon$ and $\Phi_{0,1}$ are $n$-vectors (columns) with
\begin{eqnarray}
\hat \epsilon^{i}(\omega) = \hat {\cal R}_{i1}(\omega) \hat
a_{1}(\omega), ~~\hat \Phi_{0}^{i}(\omega) =\sum_{j=1}^{n}\hat
{\cal R}_{ij}(\omega) \hat F^{j}(\omega),~~\hat
\Phi_{1}^{i}(\omega,\ \xi) =\sum_{j=1}^{n}\hat {\cal
R}_{ij}(\omega) \hat G^{j}(\omega,\ \xi)\,. \label{f2}
\end{eqnarray}
$\hat {\cal R} = \hat {\cal K}^{-1}$ and $\epsilon^{j}$ is the
energy of a hole in the $j$ sea; they are written in terms of
hyperbolic functions as
\begin{eqnarray}
&&\hat {\cal R}_{ij}(\omega) =e^{{\omega \over 2}}\ {\sinh \Big (
\min(i,j) {\omega \over 2}\Big )\, \cosh \Big (n+{1\over 2}
-\max(i,j)\Big ){\omega \over 2} \over \cosh (n+{1\over 2}){\omega
\over 2}\,\sinh{\omega \over 2} }, ~~i,j = 1, \ldots ,n
\label{RR2}
\end{eqnarray}
\begin{eqnarray}
\hat \epsilon^{j}(\omega) ={\cosh(n+{1\over 2} -j){\omega \over 2}
\over \cosh(n+{1\over 2}){\omega \over 2}}, ~~j=1,\ldots,n
\label{ener2}
\end{eqnarray}

\subsection{Scattering}

As already mentioned the main aim here is the derivation of the
exact bulk and boundary $S$-matrices. We follow the standard
formulation developed by Korepin \cite{K}, and later by Andrei and
Destri \cite{AD}. One first implement the so-called quantisation
condition,
\begin{equation}
(e^{2iNp^{l}}S-1)|\tilde \lambda_{i}^{l} \rangle = 0 \label{qc1}
\end{equation}
where $p^{l}$ is the momentum of the particle (in our case, the
hole) with rapidity $\tilde \lambda_{1}^{l}$. For the case of
$\nu$ (even) holes in $l$ sea we insert the integrated density
(\ref{sigma3}) into the quantisation condition (\ref{qc1}). We use
the dispersion relation
\begin{equation}
\epsilon^{l}(\lambda) = {1 \over 2\pi} {d \over d \lambda} \
p^{l}(\lambda)
\end{equation}
and the sum rule $\displaystyle N\int_{0}^{\tilde \lambda_{i}}
d\lambda \sigma(\lambda) \in {\bf Z}_{+}$. We end up with the
following expression for the boundary scattering amplitudes:
\begin{eqnarray}
\alpha^{+l} \alpha^{-l} = \exp \Bigl \{ 2 \pi N \int_{0}^{\tilde
\lambda_{1}}d\lambda \Bigl (\sigma^{l}(\lambda)
-2\epsilon^{l}(\lambda)\Bigr) \Bigr \}
\end{eqnarray}
with
\begin{eqnarray}
\alpha^{-l}(\lambda,\ \xi) = k_{0}(\lambda)\ k_{1}(\lambda,\ \xi),
~~\alpha^{+l}(\lambda) = k_{0}(\lambda) \label{sol1}
\end{eqnarray}
where $\alpha^{+ l}$ is realised just as the overall factor in
front of the unit matrix at the left boundary (recall that
$K^{+}=1$, whereas $K^{-}$ is given by  the solution D3).
Moreover,
\begin{eqnarray}
  k_{0}(\tilde \lambda_{1}^{l}) = \exp \Big \{ i\pi \int_{0}^{\tilde
    \lambda_{1}^{l}} d\lambda \Phi_{0}^{l}(\lambda) \Big \}, ~~k_{1}(\tilde \lambda_{1}^{l},\ \xi) = \exp \Big \{ 2i\pi \int_{0}^{\tilde
    \lambda_{1}^{l}} d\lambda \Phi_{1}^{l}(\lambda,\ \xi) \Big \},\label{ampl}
\end{eqnarray}
with $\Phi^{l}$ given by (\ref{f2}), (\ref{ef2}). We finally
restrict ourselves to $l=1$ in the first sea and we write the
latter expression in term of the Fourier transform of $\Phi^{1}$
(\ref{f2}),
\begin{eqnarray}
  k_{0}(\lambda) = \exp\Big \{ -{1\over 2} \int_{-\infty}^{\infty}
  {d\omega \over \omega} \hat \Phi_{0}^{1}(\omega)e^{-i\omega \lambda}
  \Big \}, ~~k_{1}(\lambda,\ \xi) = \exp\Big \{ - \int_{-\infty}^{\infty}
  {d\omega \over \omega} \hat \Phi_{1}^{1}(\omega,\ \xi)e^{-i\omega \lambda}
\Big \}. \label{ampl2}
\end{eqnarray}

Let us discuss first the form of the exact bulk $S$-matrix. It is
easy to compute the scattering amplitude between two holes in the
first sea. The bulk scattering amplitude comes from the
contribution of the terms of $\Phi^{1}$ given by eqs. (\ref{ef2}),
(\ref{f2}), (\ref{RR2}), with argument $\lambda \pm \tilde
\lambda_{j}$. After some algebra and using the following identity
\begin{eqnarray}
{1 \over 2}\int_{0}^{\infty}{d\omega \over
\omega}{e^{-{\mu\omega\over 2}} \over \cosh{\omega \over 2}} = \ln
\, {\Gamma({\mu+1\over 4}) \over \Gamma({\mu+3\over 4})}
\label{gamma1}
\end{eqnarray}
we conclude that the hole-hole scattering amplitude is given by
the expression
\begin{eqnarray}
S_{0}(\lambda) = {\tan \pi ({i\lambda -1\over 2n+1}) \over \tan
\pi ({i\lambda +1\over 2n+1})} \; {\Gamma({i\lambda \over 2n+1})
\over \Gamma({-i\lambda \over 2n+1})} \; {\Gamma({-i\lambda \over
2n+1}+{1\over 2}) \over \Gamma({i\lambda \over 2n+1}+{1\over 2})}
\; {\Gamma({-i\lambda + 1\over 2n+1}) \over \Gamma({i\lambda
+1\over 2n+1})} \; {\Gamma({i\lambda +1 \over 2n+1}+{1\over 2})
\over \Gamma({-i\lambda +1 \over 2n+1}+{1\over 2})}. \label{bulk1}
\end{eqnarray}
As a consistency check we compute one further eigenvalue of the
$S$-matrix. In particular if one considers the state with two
holes in the first Dirac sea, and a two-string located at the
midpoint of the two holes, the corresponding eigenvalue is given
by (see also \cite{done})
\begin{equation}
S_{b}(\lambda) = e_{1}(\lambda) S_{0}(\lambda).
\end{equation}
The explicit bulk $S$-matrix, which is a solution of the {\it
super} Yang--Baxter equation has the following structure
\begin{eqnarray}
S(\lambda) = {S_{0}(\lambda) \over (i\lambda+\kappa)(i\lambda+1)}
(\lambda(\lambda+i\kappa)1 +i(\lambda +i\kappa) P -i\lambda Q ).
\label{smatrix}
\end{eqnarray}
We now give the expressions for the boundary $S$-matrix, which
follow from (\ref{ampl}), (\ref{ampl2}), and the duplication
formula for the $\Gamma$ function
\begin{eqnarray}
2^{2x-1}\Gamma(x+{1 \over 2})\Gamma(x)= \pi^{{1\over
2}}\Gamma(2x). \label{gamma3}
\end{eqnarray}
The overall factor $k_{0}$, (\ref{ampl}), is given by
\begin{eqnarray}
k_{0}(\lambda) = Y_{0}(\lambda) {\Gamma({i\lambda \over 2n+1})
\over \Gamma({-i\lambda \over 2n+1})} \; {\Gamma({-i\lambda \over
2n+1}+{3\over 4}) \over \Gamma({i\lambda \over 2n+1}+{3\over 4})}
\; {\Gamma({i\lambda \over 2n+1}+ {1\over 2( 2n+1)}+{3\over 4})
\over \Gamma({-i\lambda\over 2n+1} +{1\over 2(2n+1)} + {3\over
4})} \; {\Gamma({-i\lambda\over 2n+1} +{1 \over 2(2n+1)}+{1\over
2}) \over \Gamma({i\lambda \over 2n+1}+{1 \over 2(2n+1)}+{1\over
2})} \label{bound1}
\end{eqnarray}
where
\begin{eqnarray}
Y_{0}(\lambda) = {\sin \pi ({i\lambda \over 2n+1} +{1\over
2(2n+1)}-{1\over 4}) \over \sin \pi ({i\lambda\over 2n+1} -{1\over
2(2n+1)}+{1\over 4})} \; {\sin \pi ({i\lambda \over 2n+1} -
{1\over 2(2n+1)}+{1\over 2}) \over \sin \pi ({i\lambda \over 2n+1}
+{1\over 2(2n+1)}-{1\over 2})} \; {\tan \pi ({i\lambda \over
2n+1}+{1\over 4(2n+1)}+{1\over 4}) \over \tan \pi ({i\lambda \over
2n+1}-{1\over 4(2n+1)}-{1\over 4})} \,. \label{Y00}
\end{eqnarray}
The $\xi$ dependent part for the D3 solution $k_{1}$, (\ref{ampl})
reads
\begin{eqnarray}
  k_{1}(\lambda,\ \xi) =  {\Gamma({i\lambda \over 2n+1} +{\xi' \over
      2n+1} +{1\over 2})\over \Gamma({-i\lambda \over 2n+1}+{\xi'
      \over 2n+1} +{1\over 2})} \  \;
  {\Gamma({-i\lambda\over 2n+1} +{\xi' \over 2n+1}) \over
    \Gamma({i\lambda \over 2n+1}+{\xi' \over 2n+1})}\ {\Gamma({-i\lambda \over
      2n+1}+{\xi' -\tilde n_{1} \over 2n+1}+{1 \over 2}) \over
    \Gamma({i\lambda \over 2n+1} + {\xi' -\tilde n_{1} \over
      2n+1}+{1 \over 2})} \;
  {\Gamma({i\lambda \over 2n+1} +{\xi' -\tilde n_{1} \over 2n+1}+1) \over
    \Gamma({-i\lambda\over 2n+1} +{\xi' -\tilde n_{1} \over 2n+1})+1)}
  \label{bound2}
\end{eqnarray}
where $\xi' =\xi -{1\over 2}$ has the fixed value found for D3, so
that the boundary $S$-matrix  satisfies the reflection equation.
Note that our solutions include the necessary CDD factors both for
the bulk and boundary matrices.

\bigskip

\textbf{Acknowledgements:} We are thankful to R.I. Nepomechie for
useful suggestions. This work was supported by the TMR Network
EUCLID: ``Integrable models and applications: from strings to
condensed matter'', contract number HPRN-CT-2002-00325.

\end{document}